# Superconductivity Favored Anisotropic Phase Stiffness in Infinite-Layer Nickelates

**Authors & Affiliations:**

Minyi Xu[1,2†], Dong Qiu[1,2†*], Minghui Xu[2†], Yehao Guo[1,2], Cheng Shen[2], Chao Yang[1], Wenjie Sun[3], Yuefeng Nie[3], Zi-Xiang Li[4,5*], Tao Xiang[4,6], Liang Qiao[1,2*], Jie Xiong[1,2*], & Yanrong Li[1]

[1]State Key Laboratory of Electronic Thin Films and Integrated Devices, University of Electronic Science and Technology of China, Chengdu, 610054, China

[2]School of Physics, University of Electronic Science and Technology of China, Chengdu, 610054, China

[3]National Laboratory of Solid State Microstructures, Jiangsu Key Laboratory of Artificial Functional Materials, College of Engineering and Applied Science and Collaborative Innovation Center of Advanced Microstructures, Nanjing University, Nanjing 210023, China

[4]Beijing National Laboratory for Condensed Matter Physics, Institute of Physics, Chinese Academy of Sciences, Beijing 100190, China

[5]School of Physical Sciences, University of Chinese Academy of Sciences, Beijing 100049, China

[6]Beijing Academy of Quantum Information Sciences, Beijing, 100193, China

† These authors contributed equally to this work.

* Corresponding authors:

dongqiu@std.uestc.edu.cn; zixiangli@iphy.ac.cn; liang.qiao@uestc.edu.cn; jiexiong@uestc.edu.cn

**Abstract**

In unconventional superconductors such as cuprates and iron pnictides and chalcogenides, phase stiffness— a measure of the energy cost associated with superconducting phase variations—is on the same order of magnitude as the strength of Cooper pairing, translating to superconductivity governed by phase fluctuations[1]. However, due to a lack of a direct experimental probe, there remains a fundamental gap in establishing microscopic picture between unconventional superconductivity and phase fluctuations. Here we show a vector current technique that allows for *in-situ* angle-resolved transport measurements, providing exclusive evidence suggesting an anisotropic nature of phase stiffness in infinite-layer nickelate superconductors. Pronounced anisotropy of in-plane resistance manifests itself in both normal and superconducting transition states, indicating crystal symmetry breaking[2-4]. Remarkably, the electric conductivity of $Nd_{0.8}Sr_{0.2}NiO_2$ peaks at 125° between the direction of the current and crystal principal axis, but this angle evolves to 160° near zero-resistance temperature. Further measurements reveal that the superconductivity is favored along a direction with minimized phase fluctuations, an orientation strikingly deviating from the symmetric direction imposed by both electronic anisotropy and the underlying crystal lattice. Identical measurements conducted on a prototypical cuprate superconductor yield consistent results, suggesting that this previously unknown behavior could be ubiquitous. By shielding insight into the contrasting anisotropy between electron fluid and superfluid, our findings provide clues for a unified framework for understanding unconventional superconductors.

## Introduction

In conventional superconductors, due to phase stiffness $J$ significantly exceeds pairing energy $\Delta$, fluctuations in the phase $\theta$ of the superconducting order parameter $\Delta e^{i\theta}$ are energetically unfavored, rendering phase fluctuations a negligible role in determining the superconducting transition temperature $T_c$[1, 5]. In contrast, in unconventional superconductors including cuprates and iron-based materials, low charge carrier density results in small superfluid density, leading to superconductivity primarily governed by the phase fluctuations. This is evidenced by the widespread presence of prominent phase fluctuations above the phase boundary of superconductivity[6, 7], and the quantitative relationship between $T_c$ and phase stiffness $J$ underscores the central role of phase fluctuations in these materials[8-10]. However, the extensive regimes adjacent to superconductivity in the phase diagrams of unconventional superconductors are filled with various intertwined orders[11, 12]. Their association with phase fluctuations complicate the understanding for a unified picture for the unconventional superconductivity.

Recently discovered superconducting nickelates provide a unique platform to address this issue. Superconductivity was first discovered in infinite-layer nickelates with $T_c$ around 9-15 K[13]. Under applied pressure, $T_c$ reached 30 K in trilayer nickelates[14] and exceeded the boiling point of liquid nitrogen in bilayer nickelates[15-17]. More recently, compressive substrate strain enhanced superconductivity with $T_c$ upto 40 K in bilayer nickelate thin films at ambient pressure[18], analogous to that in cuprate thin films. These nickelates share similar physics with cuprates[19], but notably lack translational symmetry-breaking orders at doping levels where superconductivity emerges. In particular, infinite-layer nickelates exhibit no long-range spin order[20] and charge[21-23] density-wave order near a doping level of Sr=0.2. This distinctive feature offers an unprecedented opportunity to explore interrelation between the phase fluctuations and superconductivity, isolated from other intertwined orders.

Here, we show that the establishment of superconducting phase coherence in infinite-layer nickelate superconductors $Nd_{0.8}Sr_{0.2}NiO_2$ and $La_{0.8}Sr_{0.2}NiO_2$ favors a specific direction that deviates from the symmetry imposed by both electronic nematicity and underlying lattice. Electronic nematicity in normal state is evidenced by a two-fold symmetric angular resistance, where anisotropic in-plane resistance is probed by applying current along different directions. Interestingly, resistance anisotropy persists during the superconducting transition, but with the symmetric axes of angular resistance gradually shift and ultimately stabilize at a finite value $\theta_{SC}$ as global phase coherence is achieved. Strikingly, the direction favored by electronic nematicity and superconductivity are sharply different. Further analysis of the anisotropy in non-linear current-voltage characteristics and vortex motion reveals that phase coherence is stiffer along $\theta_{SC}$ than other directions. This anisotropy in phase stiffness $J$ is fundamentally different from that of the gap energy $\Delta$ that appears in the nematic superconductivity. Consistent observations in the cuprate superconductor $La_{1.8}Sr_{0.2}CuO_4$ underscore that this unforeseen anisotropic phase stiffness may be ubiquitous among unconventional superconductors, providing crucial insights for developing a unified theoretical framework for unconventional superconductivity across various materials.

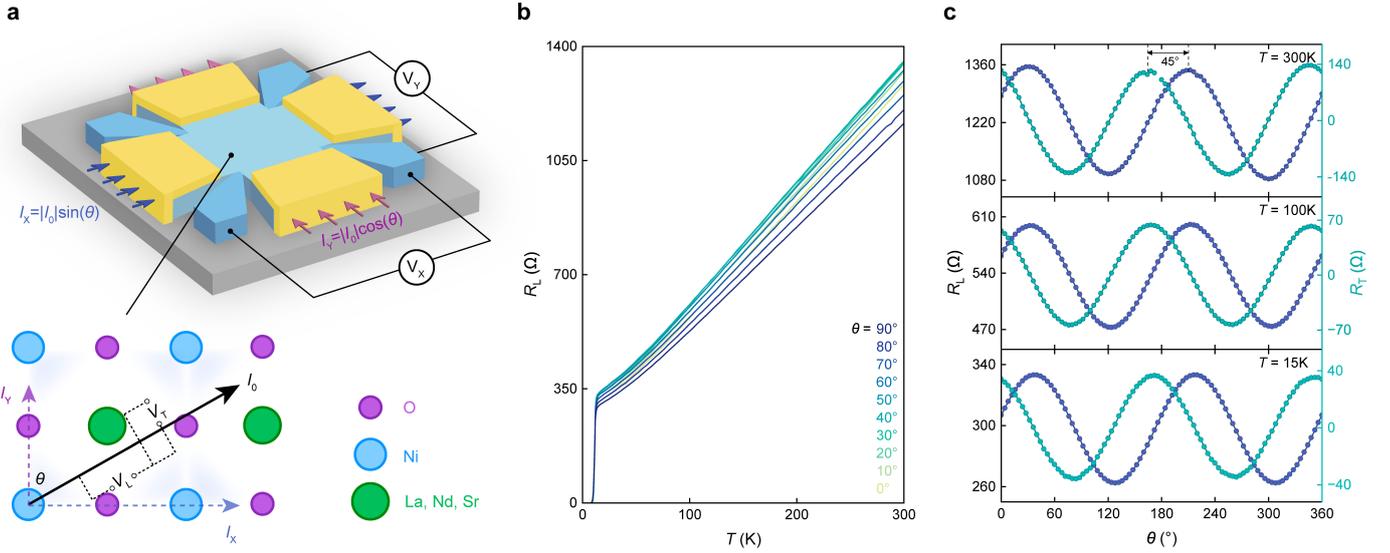

**Fig. 1| Angle-resolved measurements of in-plane resistance anisotropy in infinite-layer superconducting nickelate films.** **a,** Illustration of the 'vector current' technique. The net current $I_0$ is composed of two orthogonal injected DC current, $I_X$ (blue arrow) and $I_Y$ (purple arrow), where the angle $\theta$ between the net current $I_0$ and crystal principal axis as well as the amplitude of $I_0$ can be tuned continuously. **b,** Temperature dependence of longitudinal resistance $R_L$ of $Nd_{0.8}Sr_{0.2}NiO_2$ at selected $\theta$ ranging from 0° to 90°. **c,** Angle-dependence of longitudinal resistance $R_L(\theta)$ (blue line) and transverse resistance $R_T(\theta)$ (green line) of $Nd_{0.8}Sr_{0.2}NiO_2$ at $T$ = 300 K, 100 K and 15 K. Black arrow marks the 45° phase shift between $R_L(\theta)$ and $R_T(\theta)$.

## Results

### Electronic nematicity in normal state

$Nd_{0.8}Sr_{0.2}NiO_2$ films with the thickness of 15 nm were grown on tetragonal $SrTiO_3$ substrates by pulsed laser deposition with successive topotactic reduction[24] (see Methods). The residual resistance ratio is 4.3 and superconducting onset temperature $T_c^{onset}$ of the film is 15 K (temperature at which the $R_L(T)$ deviates from the linear extrapolation of the normal state). Structural characterization indicates that the lattice symmetry of $Nd_{0.8}Sr_{0.2}NiO_2$ is tetragonal without observable lattice distortions (see Methods). These observations are consistent with the literature[19]. We utilize the 'vector current' technique (see Fig. 1a) to probe the angular resistance of tetragonal superconducting films[25-28]. Due to the vector nature of the electric field, the direction and amplitude of the net current $I_0$ can be continuously tuned by two orthogonal DC currents, $I_X = I_0 \sin\theta$ and $I_Y = I_0 \cos\theta$ ($\theta$ is the angle between $I_0$ and crystal principal axis, see Fig. 1a). As a result, the corresponding longitudinal and transverse resistance along the $I_0(\theta)$ can be measured as $R_L = (V_X \sin\theta + V_Y \cos\theta)/I_0$ and $R_T = (V_X \cos\theta - V_Y \sin\theta)/I_0$, respectively. The current direction and amplitude can be arbitrarily tuned *in-situ* in a single device, enabling the detection of possible anisotropy of the physical properties. The validity and reliability of the 'vector current' method are confirmed by COMSOL simulations and control experiments, as shown in the Supplementary Information and Extended Data Fig. 2.

Fig. 1b depicts the temperature dependence of the longitudinal resistance $R_L(T)$ at representative directions for $Nd_{0.8}Sr_{0.2}NiO_2$, showing a distinguishable resistance anisotropy between different directions. The angle dependence of the $R_L(\theta)$ and $R_T(\theta)$ at representative temperatures is illustrated in Fig. 1c, revealing a clear oscillation with a 180° period, observed from room temperature down to $T_c^{onset}$. Both $R_L(\theta)$ and $R_T(\theta)$ display the same periods and amplitudes, but are phase-shifted by 45°. Thus, the angular resistance can be described by Equation (1) and (2):

$$R_L(\theta) = R_{L0} - \Delta R \cos[2(\theta - \theta_0)] \tag{1}$$

$$R_T(\theta) = -\Delta R \sin[2(\theta - \theta_0)] \tag{2}$$

where $R_{L0}$ is the isotropic background of the angular resistance, $\Delta R$ is the anisotropic resistance amplitude, and $\theta_0$ represents the direction of the resistance minimum. The $\sin(2\theta)$ relationship between resistance and current direction $\theta$ is a natural result when the current is not aligned with the principal axes of the two-fold symmetric resistance[2, 4, 29] (Supplementary Information Section 2). Control experiments, including angular resistance measurements on a 'ring-shaped' device and on various materials (gold, anoxic $SrTiO_3$, and NbN), were conducted to ensure the two-fold symmetric resistance is intrinsic to the $Nd_{0.8}Sr_{0.2}NiO_2$ film (Extended Data Fig. 2–3). This two-fold symmetry in angular resistance indicates that the rotational symmetry of the tetragonal $Nd_{0.8}Sr_{0.2}NiO_2$ thin film is broken. Since the translational symmetry in the infinite-layer superconducting nickelate is preserved due to the absence of the long-range spin[20] and charge[21-23] density-wave order, we attribute the two-fold symmetric angular resistance to electronic nematicity. Additionally, the direction of lowest resistance in the normal state appears at $\theta_N = 125°$, deviating from the crystal lattice's principal axes (0° or 90°), suggesting the coexistence of $B_{1g}$ and $B_{2g}$ nematic orders. The persistence of resistance anisotropy up to room temperature (Fig. 1c) resembles that observed in $La_{2-x}Sr_xCuO_4$[2, 4, 30] and $Sr_2RuO_4$[31], awaiting further exploration.

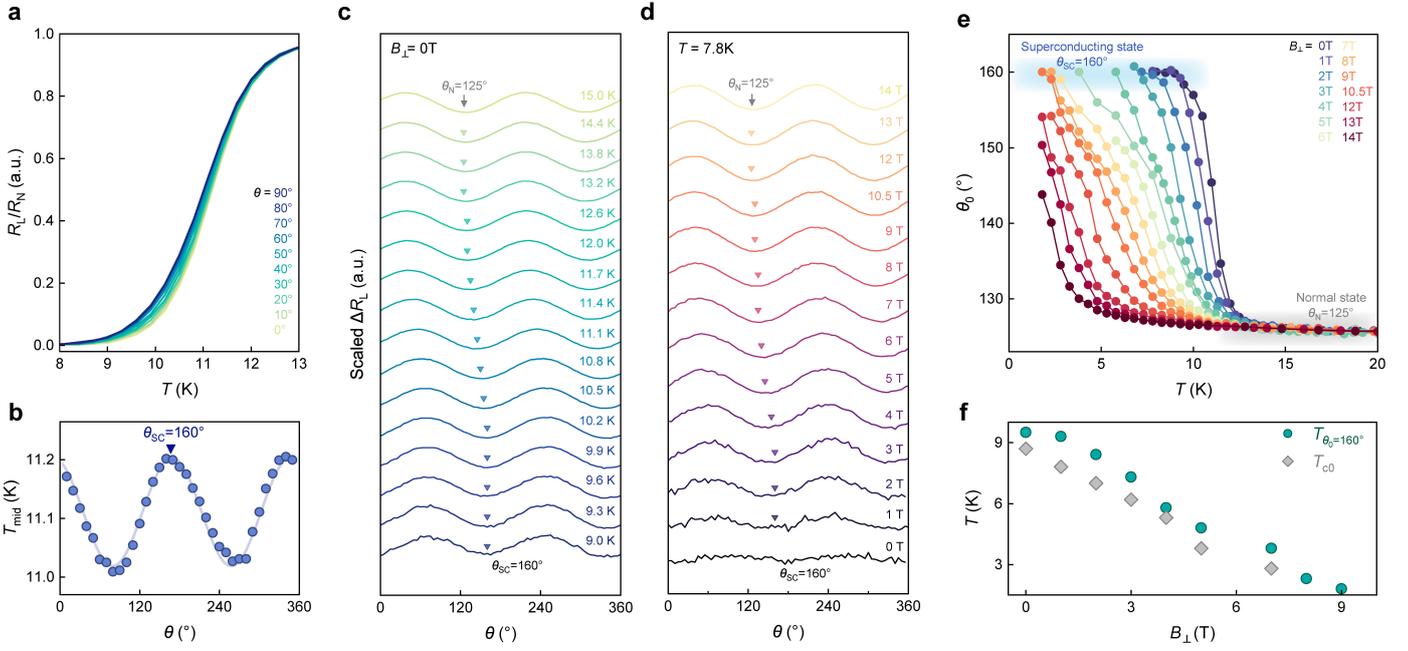

**Fig. 2| Shift of the symmetric axis $\theta_0$ during the superconducting transition. a,** Superconducting transition of Nd$_{0.8}$Sr$_{0.2}$NiO$_2$ at representative current direction $\theta$, where $R_N$ is defined as the resistance at $T$=15 K. **b,** Anisotropic superconducting transition of Nd$_{0.8}$Sr$_{0.2}$NiO$_2$, where $T_{mid}$ is defined as the temperature where $R_L$ reaches the 50% of the $R_N$. Arrow marks the maximum of $T_{mid}$ the locates at $\theta_{SC}$=160°. **c,** Angle-dependent scaled $\Delta R_L(\theta)$ at a series of temperature $T$ under zero field. **d,** Angle-dependent scaled $\Delta R_L(\theta)$ measured under a series of perpendicular magnetic fields $B_\perp$ ranges from 0T to 14T, at fixed temperature $T$=7.8K. Curves in **c-d** are scaled and vertically shifted for comparison (original data are presented in Extended Data Fig. 4). Triangles denotes the location of the lowest longitudinal resistance as the $\theta_0$. **e,** The symmetric axis $\theta_0$ of $\Delta R_L(\theta)$ as a function of temperature $T$ and perpendicular magnetic field $B_\perp$, where the shaded area denotes the $\theta_N$ and $\theta_{SC}$. **f,** Relevance between the zero-resistance temperature $T_{c0}$ and temperature where $\theta_0$ reaches 160°, $T_{\theta_0=160°}$. The $T_{c0}$ denotes the zero-resistance temperature when $R_{L0}(T)$ reaches 0.5% $R_{L0}|_{T=15K}$.

## Dichotomy of anisotropy in normal state and superconductivity

Further cooling below $T_c^{onset}$ reveals directional anisotropy in the sharpness of the superconducting transition (Fig. 2a). This anisotropy is quantified by defining $T_{mid}$ at a given direction $\theta$ as the temperature where resistance reaches 50% $R_N$. Fig. 2b shows that $T_{mid}$ is clearly two-fold symmetric, with the maximum at $\theta_{SC}$=160°. The $\theta_{SC}$ consistently aligns at 160° in the analysis of the anisotropy of temperature when resistance reaches 30% and 90% $R_N$ (Fig. S5). The deviation of $\theta_{SC}$ from $\theta_N$ suggests that the anisotropy of superconductivity is different from that in normal state.

We investigate the evolution of the symmetric axes $\theta_0$ during the superconducting transition. Fig. 2c shows the temperature dependence of the anisotropic part of the angular resistance $\Delta R_L(\theta) = R_L(\theta) - R_{L0}$. At high temperatures, the $\theta_0$ (marked by triangles in Fig. 2c) for the normal state is fixed at $\theta_N$=125°. However, a different picture appears below the $T_c^{onset}$: $\theta_0$ starts to shift and eventually saturate at $\theta_{SC}$ =160° near the zero-resistance temperature $T_{c0}$. As the superconductivity is suppressed by the perpendicular magnetic field $B_\perp$, $\theta_0$ shifts back to the normal state value $\theta_N$ (Fig. 2d). As shown in Fig. 2e, $\theta_0$ switches between the $\theta_N$ for the normal state and $\theta_{SC}$ for the superconducting state. This non-trivial dependence of $\theta_0$ on temperature and perpendicular magnetic field in $Nd_{0.8}Sr_{0.2}NiO_2$ film provides strong evidence against the two-fold symmetric resistance caused by sample inhomogeneity or macroscopic lattice distortion. We interpret the distinct $\theta_N$ and $\theta_{SC}$ as evidence of the dichotomy of the anisotropy in normal state and superconducting state.

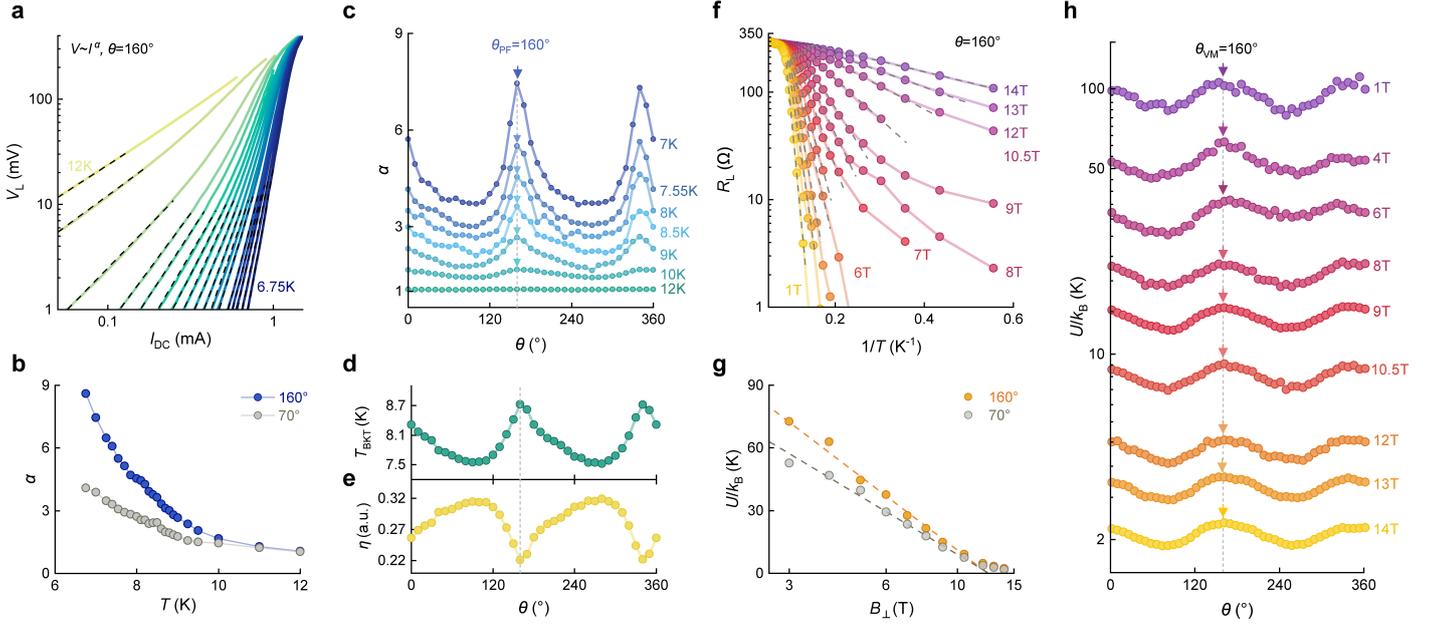

**Fig. 3| Anisotropy in superconducting phase fluctuations and vortex motion. a,** Representative current-voltage characteristic measured at $\theta$=160° for temperature ranges from 6.75 K to 12 K. Dashed black lines show the fit with the power-law relation given by $V \sim I^\alpha$. **b,** Extracted $\alpha$ as a function of temperature for $\theta$=70° and 160°. **c,** Angle-dependent $\alpha(\theta)$ at representative temperatures, arrows mark the maximum locates at $\theta_{PF}$=160°. **d,** Angle-dependent $T_{BKT}(\theta)$ as obtained by $\alpha(\theta)$=3, arrow marks the maximum locates at $\theta_{PF}$=160°. **e,** Angle-dependent $\eta(\theta)$ as the intensity of phase fluctuations, arrow marks the minimum locates at $\theta_{PF}$=160°. **f,** Arrhenius plots of temperature-dependent resistance $R_L(T)$ at $\theta$=160° under a series of perpendicular magnetic field $B_\perp$. Dashed lines are fitted by the $R_L(T) = R_0 \exp(-U(B)/k_B T)$. **g,** Extracted activation energy $U/k_B$ as a function of perpendicular magnetic field for $\theta$=70° and 160°. Dashed lines indicate the $U/k_B$ is proportional to $\ln(B^{-1})$. **h,** Angle-dependent $U(\theta)/k_B$ at various perpendicular magnetic field $B_\perp$, arrows mark the maximum locates at $\theta_{VM}$=160°.

## Anisotropic superconducting phase stiffness

Fig. 2f shows that temperature where $\theta_0$ reaches $\theta_{SC}$ coincides with $T_{c0}$ at each magnetic field, suggesting $\theta_{SC}$ serve as an indicator for global superconducting phase coherence. To understand how the phase coherence develops, we performed current-voltage measurements at different current directions $\theta$ to investigate superconducting phase fluctuations. Near the zero-resistance temperature, thermal fluctuations soften vortex-antivortex pairs, making them more susceptible to current-induced unbinding, leading to non-linear I-V characteristics, $V \sim I^{\alpha(T)}$ (Fig. 3a). Fig. 3b shows the extracted exponent $\alpha(T)$ at representative directions $\theta$ =70° and 160°, where $\alpha$ develops pronounced anisotropy below 12 K. The maximum $\alpha(\theta)$ locates at $\theta_{PF}$=160° (Fig. 3c), indicating the vortex-antivortex pairs are most resistant to current-unbinding at 160°. This anisotropy in the current unbinding process results in an anisotropic establishment of global phase coherence. By adopting the conventional notion of the Berezinskii-Kosterlitz-Thouless (BKT)transition[32, 33], the anisotropic establishment of global phase coherence manifests as the BKT transition temperature $T_{BKT}$ (temperature when $\alpha$ =3), as shown in Fig. 3d. This is the evidence of anisotropic phase fluctuations, the intensity of which can be quantified by the normalized superconducting transition width, $\eta = (T_{mid} - T_{BKT})/T_{mid}$. Fig. 3e indicates that phase fluctuations are anisotropic, in which the phase fluctuations along 160° is weaker than other directions.

The key to understanding anisotropic superconducting phase fluctuations lies in vortex dynamics. In type-II superconductors, vortex motion leads to phase decoherence[34]. Under a perpendicular magnetic field, vortices are generated and move with the activation of thermal fluctuations, described as 'thermally-assisted flux-flow' (TAFF)[34]. As shown in Fig. 3f, the temperature-dependent resistance follows the TAFF behavior with $R_L(T) = R_0 \exp(-U(B)/k_B T)$, where $k_B$ is Boltzmann's constant and $U(B)$ is the magnetic field-dependent activation energy for vortex[35]. The $U \propto \ln(B)$ relation in Fig. 3g are consistent with the thermally assisted collective vortex-creep model[35]. Fig. 3h shows that $U$ exhibits two-fold symmetry under different magnetic field up to 14 T. The largest vortex activation energy occurs at $\theta_{VM}$=160°, suggesting vortex motion are more difficult to activate in this direction. Thus, superconducting phase coherence is stiffer along 160°.

We further investigate the anisotropy of superconducting state by extracting superfluid phase stiffness $J$. Based on the model of thermally assisted collective vortex-creep[35], the slope of $U \propto \ln(B)$ in Fig. 3g relates to the penetration depth (See Methods) thus the phase stiffness $J$ with the following ration

$$J = 16 \cdot \frac{a}{t} \cdot \frac{d(U/k_B)}{d\ln(B)} \qquad (3)$$

where $a$ is the separation length between adjacent superconducting planes for layered materials[1, 5], $t$ is the sample thickness. We extract the phase stiffness along different directions, Fig. 4a shows the angle-dependent phase stiffness $J(\theta)$ for $Nd_{0.8}Sr_{0.2}NiO_2$ in a polar plot. $J(\theta)$ exhibits clear two-fold symmetry with its maximum at $\theta_{SC}$=160°, highlighting the anisotropy in the phase stiffness. Identical measurements and analysis were performed on La$_{0.8}$Sr$_{0.2}$NiO$_2$ and La$_{1.8}$Sr$_{0.2}$CuO$_4$ films to testify whether the anisotropic phase stiffness is unique to Nd$_{0.8}$Sr$_{0.2}$NiO$_2$. Fig. 4b and 4c present the angle-dependent phase stiffness $J(\theta)$ for La$_{0.8}$Sr$_{0.2}$NiO$_2$ and La$_{1.8}$Sr$_{0.2}$CuO$_4$ in polar plots. In all three materials, $J(\theta)$ displays two-fold symmetry, with its maximum located at $\theta_{SC}$, aligning with the direction where $T_{mid}$ is maximized. The phase stiffness to transition temperature ratio, $J/T_c$ is approximately 1 in all three materials, considerably lower than in the conventional superconductors[1, 5], indicating the superconductivity is controlled by the phase stiffness. This conclusion is further supported by the alignment of $\theta_{SC}$, $\theta_{PF}$ and $\theta_{VM}$ in all three materials (Extended Data Fig. 7), demonstrating that the symmetry of superconducting properties is consistently controlled by the anisotropic phase stiffness. Since $\theta_{SC}$ is decoupled from $\theta_N$ imposed by electronic nematicity or 0°/90° of the crystal principal axes in all three materials, we infer that superconducting phase coherence is not preferentially favored along the symmetric direction imposed by electronic anisotropy. Although statistics shows that exact values for $\theta_N$ and $\theta_{SC}$ fluctuate in different samples within a considerable range (Extended Data Table. 2), the main conclusion—the dichotomy of anisotropy in normal state and superconducting state—remains highly reproducible.

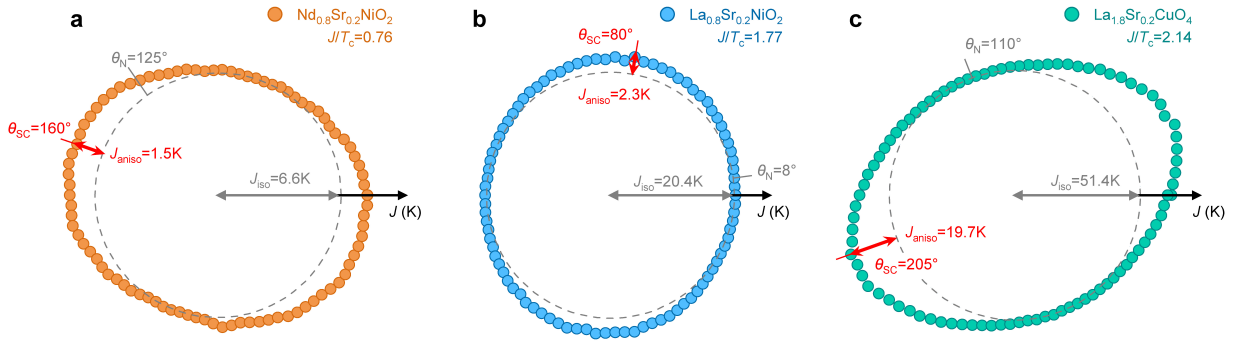

**Fig. 4| Anisotropic phase stiffness in three different unconventional superconductors. a,** Linear-scale polar plot of the angle-dependent $J(\theta)$ for Nd$_{0.8}$Sr$_{0.2}$NiO$_2$. The isotropic background of the phase stiffness $J_{iso}$ is defined as the minimum of $J(\theta)$, shown as the dashed grey line. The difference between the maximum and minimum of the anisotropic $J(\theta)$ denotes the $J_{aniso}$, shown as the red arrow. The $\theta_N$ and $\theta_{SC}$ is defined as the location of $R_L(\theta)$ minimum measured at $T_c^{onset}$ and $T_{c0}$, respectively. The $\theta_{SC}$ coincides with direction where the maximum of $J(\theta)$ occurs in all three materials. The $\theta_N$ and $\theta_{SC}$ is marked by the grey and red lines, respectively. **b-c,** Linear-scale polar plot of the angle-dependent $J(\theta)$ for La$_{0.8}$Sr$_{0.2}$NiO$_2$ and La$_{1.8}$Sr$_{0.2}$CuO$_4$. Markers are defined as the same as that for Nd$_{0.8}$Sr$_{0.2}$NiO$_2$. $J/T_c$ is calculated by using $J_{iso}$ and $T_{c0}$ to represent the phase stiffness and superconducting phase coherence temperature. In three materials, $J/T_c$ is found to be close to 1. this result is also consistent by choosing maximized phase stiffness $J_{aniso} + J_{iso}$ as $J$.

## Discussion

The present work uncovers an exotic superconductivity with anisotropic phase stiffness. While previous studies have primarily examined the role of nematic fluctuations in Cooper pair formation, our observations provide a unique insight into addressing their complexing interrelation by highlighting a connection between nematicity and the emergence of phase coherence, the critical importance of which is a defining feature of unconventional superconductor. The consistent observation of this misaligned anisotropy in both nickel- and copper-based superconductors suggests a fundamental but unforeseen mechanism governing the nematicity and superconductivity. Moreover, similar phenomena potentially exist in diverse quantum materials (Extended Data Table. 3), implying that the misaligned anisotropy of normal and superconducting states is an essential and universal property. However, this misalignment has been largely overlooked in previous research, and its underlying origins remain poorly understood.

Building on our experimental data, we seek to discuss possible origins for the observations. Firstly, the shift of the symmetric axis of the anisotropic resistance is highly reproducible across different measurement setups—both in 'vector current' and 'ring-shaped' devices—and across various samples (See Methods). This consistency allows us to exclude macroscopic orthogonal distortions of the crystal, which would typically pin the symmetric axis along the crystal axes[36]. Secondly, the absence of the long-range spin or charge order in both superconducting (Nd,Sr)NiO$_2$[20, 21] and (La,Sr)NiO$_2$[22] renders the magnetism[37] and charge density-wave irrelevant. Thirdly, the consistency of observations between nickelates and cuprates (Extended Data Fig. 7) suggests that features specific to nickelate thin films—such as topotactic reduction[23, 38], Nd$^{3+}$ $4f$ magnetic moments[39], Ruddlesden-Popper stacking faults and chemical inhomogeneity[40]—do not critically impact our findings. Moreover, the consistent behavior across all three materials points to an intrinsic anisotropy in phase stiffness, distinct from superconducting stripes typically induced by coupling with an external ferromagnetic overlayer in heterostructures[37, 41]. It's worth further exploring other possible origins from the properties of complex oxide thin films, such as orbital ordering driven by surface reconstructions[42] or the Jahn-Teller effect[43], or the dynamic orthorhombic distortion at atomic scale[44].

Our observations of the anisotropic phase stiffness can hardly reconcile with the current understanding. By comparing our observation with recent measurements of angle-resolved magnetoresistance (ADMR) on infinite-layer nickelates[39, 45], we demonstrate anisotropy of phase stiffness appears independent of the symmetry of the superconducting order parameter. Instead of characterizing phase stiffness, ADMR measurements reveal gap symmetry by applying an in-plane magnetic field $B_{\parallel}$ to break Cooper pairs. In this way, although the rotational symmetry is preserved in Nd$_{0.8}$Sr$_{0.2}$NiO$_2$ near zero field[45], that of the phase stiffness is found to be broken in our measurements (Extended Data Fig. 7). This indicates that the superconducting state with anisotropic phase stiffness we uncover is essentially distinct from the nematic superconductivity manifested by the anisotropic order parameter[46]. Furthermore, the observed anisotropic phase stiffness does not arise from the coupling between superconductivity and unidirectional (charge or spin) density-wave order, which creates a secondary spatially-modulated pair-density-wave (PDW) order, naturally leading to anisotropic phase stiffness[47]. The lack of density-wave order in the normal state of infinite-layer nickelate superconductors rules out this scenario. Nonetheless, whether a primary PDW state underlies the anisotropy in phase stiffness remains an open and intriguing question for future study[48].

Thus far, we suggest the observed superconductivity with anisotropic phase stiffness unifies two critical ingredients—phase fluctuations and nematicity—through an unforeseen and potentially universal mechanism. The misaligned anisotropy in electronic fluid and superfluid suggests that the relationship between nematicity and superconductivity extend beyond mere competition[29] and cooperation[49, 50], providing new guidance for understanding them within a unified framework. Broadly, angle-resolved transport measurements provide valuable tools for tracking quantitative evolution of superconducting phase coherence along specific directions, stimulating the theoretical and numerical investigations of the quantum many-body models to further address the complex interrelation. Particularly, recent numerical calculation on the Hubbard model suggests the coexistence of superconductivity and partially filled stripe order in certain parameter regime[51], indicating a possible microscopic origin for anisotropic phase stiffness. Further exploration of directional phase stiffness on the numerical level will promote fathoming the fundamental properties of the unconventional superconductivity and associated intertwined electronic orders. Our work motivates similar measurements on other unconventional superconductors, potentially paving the way toward the universal mechanism of unconventional superconductivity.

**Acknowledgments**

The authors would like to thank Yanwu Xie, Kun Jiang, Chunyu Guo, and Haiwen Liu for fruitful discussions. This work was supported by the National Key Research and Development Program of China (2021YFA0718800, 2023YFA1406301), the National Natural Science Foundation of China (52021001, 12274061, 12347107, 12474146).

**Author Contributions**

M.Y.X., J.X. and L.Q. conceived the project. D.Q. and Z.X.L. supervised the project. D.Q. and M.Y.X. designed the angular resistance measurement device. M.H.X. and L.Q. synthesized the $Nd_{0.8}Sr_{0.2}NiO_2$ films. G.Y.H performed COMSOL simulations. W.J.S. and Y.F.N. synthesized the $La_{0.8}Sr_{0.2}NiO_2$ films. M.Y.X. performed the angular resistance measurement with help from D.Q.. M.H.X. performed structure characterization with help from L.Q.. D.Q. and M.Y.X. analyzed the experimental data. Z.X.L. contributed to the data analysis and theoretical discussions. M.Y.X., D.Q., Z.X.L., L.Q., and J.X. wrote the manuscript with input from T.X., C.S., C.Y., and Y.R.L.

**Competing Interests**

The authors declare no competing interests.

**Data availability**

Source data are provided with this paper. All other data that support the findings of this study are available from the corresponding authors upon reasonable request.

## Methods

### Films growth and structure characterization

Perovskite $Nd_{0.8}Sr_{0.2}NiO_3$ films with the thickness of 15 nm, were grown on single-crystal $SrTiO_3$ (001) substrates using pulsed laser deposition (PLD, Demcon TSST) with a KrF excimer laser ($\lambda$=248 nm). The $SrTiO_3$ (001) substrates were pre-treated by HF etching and annealed at 1000 °C for 2 hours to obtain atomically flat $TiO_2$-terminated surfaces. A uniform laser spots (1.2 mm × 2.8 mm) was used for ablation, generated by aperture imaging. The film growth temperature was set at 580°C, with the oxygen partial pressure maintained at 200 mTorr. The topological-oriented reduction was conducted under the optimized conditions established in our previous study[52].

High-angle annular dark field scanning transmission electron microscopy (HAADF-STEM) was used to examine structural transformations following the reduction process (Extended Data Fig. 1a). The $Nd_{0.8}Sr_{0.2}NiO_2$ thin films primarily consists of the infinite planar layered structure with no significant blockiness or planar faults. Additionally, the Energy-dispersive X-ray spectroscopy (EDS) was also performed on superconducting samples to assess element distribution and composition. Extended Data Fig. 1b shows sharp contrast between the film and substrate, indicating high crystal quality. EDS analysis confirmed the film composition closely matches the nominal $Nd_{0.8}Sr_{0.2}NiO_2$ formula. Reciprocal space mapping (RSM) near the $SrTiO_3$(103) reflection provided further structural details (Extended Data Fig 1c), showing that the film remained strained to the tetragonal $SrTiO_3$ lattice, indicating the in-plane lattice symmetry of the film is consistent with that of the $SrTiO_3$. X-ray diffraction results (Extended Data Fig. 1d) revealed a decrement in the $c$-axis lattice constant from 3.76 Å to 3.36 Å after reduction process. The X-ray reflectivity (XRR) analysis (inset of Extended Data Fig. 1d) confirmed the post-reduction $Nd_{0.8}Sr_{0.2}NiO_2$ film thickness to be 14.6 nm, in agreement with the nominal thickness of 15 nm film.

### Angular resistance measurements

Electrodes (20 nm silver/20 nm gold) were evaporated onto the films, ensuring contact resistance below 1 Ω. For the 'vector current' device, the films were patterned into an eight-terminal configuration (Fig. 1a) via standard photolithography. The central area of the device was 100 μm × 100 μm. DC currents were applied using Keithley Model 6221 current sources, and DC voltages were measured with a Keithley Model 2182A nanovoltmeter. A 'pulse-delta' method was used, where the current source generated alternating positive ($+I$) and negative ($-I$) pulses, and the nanovoltmeter averaged the voltages as $[V(+I) - V(-I)]/2$. This method eliminates the influence of (1) the drift voltage of the nanovoltmeter and (2) Joule heating in the measurement of current-voltage characteristics. For the ring-shaped device (Extended Data Fig. 3), angular resistance is characterized by measuring the transverse resistance with the standard lock-in technique. The diameter of the measured area is 1 mm. The temperature and magnetic field dependence of resistivity was measured using commercial cryostat.

**Validity of 'vector current' method**

To rule out artifacts or extrinsic factors that could contribute to the observed two-fold symmetric resistance, identical angular measurements were conducted on gold films, anoxic SrTiO$_3$ films, and NbN films Isotropic resistance observed in the gold films (Extended Data Fig. 2), as supported by COMSOL simulations (Fig. S5), confirmed that the vector current method does not introduce spurious two-fold symmetric resistance. The isotropic resistance in anoxic SrTiO$_3$ films indicates that oxygen vacancy distribution alone does not induce broken rotational symmetry. Additionally, the isotropic resistance in conventional $s$-wave NbN superconductors implies that the two-fold symmetric angular resistance is specific to unconventional superconductors.

To further assess the influence of structural disorder, simulations were conducted, revealing that macroscopic inhomogeneity (e.g., scratches, interface particles) can induce a background signal $R_{T0}$ in transverse resistance measurements (Fig. S6). However, given the negligible $R_{T0}$ in our measurements (Fig. 1c, Extended Data Fig.3), we exclude macroscopic structural disorder as a significant contributor to the observed resistance anisotropy in La$_{0.8}$Sr$_{0.2}$NiO$_2$, Nd$_{0.8}$Sr$_{0.2}$NiO$_2$ and La$_{1.8}$Sr$_{0.2}$CuO$_4$.

**Reproducibility and statistics**

The central observation of our work is the shift of symmetric axis during the superconducting transition, which indicates the dichotomy of anisotropy in normal state and superconducting state. Our key conclusion is that anisotropy of the superconductivity originates from anisotropic phase stiffness. To validate these findings, we conducted identical measurements on six devices, ensuring (1) reproducibility of the central observation and (2) consistency of the key conclusion across different sample batches.

Extended Data Table 1 summarizes the symmetric axes of normal state ($\theta_N$) and superconducting state ($\theta_{SC}$) for six devices. We found that the 'dichotomy of anisotropy' is highly reproducible, with $\theta_N$ and $\theta_{SC}$ fluctuating within a narrow range. This variation may be attributed to either (1) discrepancies between $\theta = 0°$ and the principal axis of the film or (2) variations in thin film quality across different sample batches (Extended Data Fig. 10). We note that the $\theta_N$ for La$_{1.8}$Sr$_{0.2}$CuO$_4$ (p=0.20) in our measurements is consistent with $\theta_N$ for La$_{1.8}$Sr$_{0.2}$CuO$_4$ (p=0.21) measured previously by other group[2], reinforcing the reproducibility of our measurements.

Identical measurements were also performed on NSNO-S3, LSNO-S1, and LSCO-S3. Extended Data Fig. 5-7 demonstrate that the 'dichotomy of anisotropy' is present across these three materials, exhibiting similar behaviors. Additionally, we measured non-linear I-V characteristics and thermally-assisted flux-flow, along with corresponding analyses for La$_{0.8}$Sr$_{0.2}$NiO$_2$ and La$_{1.8}$Sr$_{0.2}$CuO$_4$ (Supplementary Section 4). All physical quantities related to superconductivity were consistently aligned among Nd$_{0.8}$Sr$_{0.2}$NiO$_2$, La$_{0.8}$Sr$_{0.2}$NiO$_2$ and La$_{1.8}$Sr$_{0.2}$CuO$_4$ (Extended Data Fig. 7), supporting the conclusion that 'anisotropy in superconductivity originates from the anisotropic phase stiffness'.

**Analysis of anisotropy of the superconducting phase stiffness**

In this section, we discuss the details of analysis on the anisotropic phase stiffness. The anisotropy of superconducting phase fluctuations (Fig. 3c) and vortex activation energy (Fig. 3h) implies the superconducting phase stiffness can be anisotropic. However, conventional characterizations of superconducting phase stiffness involve measuring penetration depth, which reflects an average of the in-plane properties $\lambda_{ab}$ and the possible difference between the $\lambda_a$ and $\lambda_b$ is hidden. Besides, there's limited theory focusing on the in-plane anisotropy of Cooper-pair mass/penetration depth/phase stiffness[30]. Hence, the anisotropy of phase stiffness is hard to reveal due to limited tools. Our strategy at current stage is to use different theoretical models to extract phase stiffness, and evaluate (1) whether the value of the extracted phase stiffness is consistent with that measured in experiments, and (2) whether different models yield consistency of anisotropic phase stiffness.

We firstly analyze based on the Feigelman-Geshkenbein-Larkin (FGL) model[35], which is applicable for various layered materials with large anisotropy. Fig. 3g depicts that the activation energy of vortex motion exhibits a linear dependence on perpendicular magnetic fields in a semilogarithmic plot, $U(B) \propto \ln(B)$, a behavior observed in many 2D superconductors[53]. This is also consistent with the two-dimensional nature of superconductivity in Nd$_{0.8}$Sr$_{0.2}$NiO$_2$[45] and La$_{0.8}$Sr$_{0.2}$NiO$_2$[54] thin film. According to the FGL model, the slope of $U(B) - \ln(B)$ is proportional to the penetration depth $\lambda^{-2}$

$$U(B) = \frac{\Phi_0^2 t}{64\mu_0 \pi^2 \lambda^2} \ln\left(\frac{B_0}{B}\right) \tag{4}$$

where $t$ is the sample thickness, $\Phi_0$ is the magnetic flux quantum, $\mu_0$ is the vacuum permittivity, $\lambda$ is the penetration depth, $B_0$ is the numeric parameter, $k_B$ is the Boltzmann constant, $e$ is the element of charge. By recalling the phase stiffness $J = \hbar^2 a / 4\mu_0 e^2 \lambda^2$, we connect the slope of $U \propto \ln(B)$ and phase stiffness by:

$$J = \frac{a}{t} \cdot \frac{16\pi^2 \hbar^2}{e^2 \Phi_0^2} \cdot \frac{dU/k_B}{d\ln(B)} \tag{5}$$

where $a$ is the separation length between adjacent superconducting plane for layered materials[1, 5], $\hbar$ is the reduced Planck constant. Eq. (3) shown in the main-text is obtained by inserting the definition of flux quantum $\Phi_0 = \pi\hbar/e$ into Eq. (5). By considering there's two superconducting planes in one unit-cell, the $a$ should be half of the $c$-axis constant, measured as 0.168 nm, 0.173 nm, and 0.66 nm for $Nd_{0.8}Sr_{0.2}NiO_2$, $La_{0.8}Sr_{0.2}NiO_2$, and $La_{1.8}Sr_{0.2}CuO_4$, respectively. As shown in Extended Data Table 2, an excellent consistence is found between the extracted $J$ and the measured zero-temperature phase stiffness $J_0$ in $La_{1.8}Sr_{0.2}CuO_4$[9]. Although the TGL-model describe the properties at finite-temperature, it provides reasonable value of $J$ which is close to the zero-temperature phase stiffness $J_0$, as revealed by previous study[55]. The accurate value of $\lambda_0$ for As for the superconducting infinite-layer nickelates has not been experimentally confirmed yet[19], awaiting further experimental confirmation. We further use this analysis to extract $J$ in other materials, we find the consistency between the extracted $J$ and experimentally measured zero-temperature $J_0$ (Extended Data Table 2). This demonstrates that, the value of the extracted phase stiffness in our experiments is reasonable.

Below we discuss the anisotropy of the phase stiffness analyzed by different strategies. Firstly, two separated theoretical models connect the phase stiffness to the vortex activation energy[35, 56]. In both models, the slope of $U \propto \ln(B)$ is proportional to $J$ through the relation of $d(U/k_B)/d\ln(B) \propto \lambda^{-2} \propto J$. Therefore, we can measure the $d(U/k_B)/d\ln(B)$ along different $\theta$ to extract anisotropic phase stiffness $J$. Besides, the exponent $\alpha$ in the non-linear I-V curve $V \sim I^\alpha$ is proportional to the phase stiffness through the relation of $\alpha = 1 + \pi J / T$. Furthermore, by adopting the conventional notion of BKT transition, the BKT transition temperature is directly proportional to the phase stiffness (Fig. S6), $T_{\text{BKT}} = \pi J / 2k_B$. To eliminate the influence of theoretical details on the analysis, we choose the normalized phase stiffness $\Delta J / J_{\text{iso}}$ for estimating anisotropy. Extended Data Fig. 8 shows the phase stiffness $J$ extracted from (a) vortex activation energy, (b) non-linear I-V curve and (c) BKT transition exhibits a clear two-fold symmetry, with the symmetric axis consistently align with $\theta_{SC}$. This consistency strongly supports the existence of anisotropic superconducting phase stiffness in $Nd_{0.8}Sr_{0.2}NiO_2$, $La_{0.8}Sr_{0.2}NiO_2$ and $La_{1.8}Sr_{0.2}CuO_4$.

**Hints of the dichotomy of anisotropy in other materials**

In our study, we define evidence for the "dichotomy of anisotropy" as the distinct symmetry axes in the normal and superconducting states, manifesting as a shift in the symmetric axis of angular resistance, $R_L(\theta)$, during the superconducting transition (Fig. 2c-2d). If we extend this concept to include the shift of the symmetric axis of a physical quantity related to superconductivity with external parameters (e.g. temperature, magnetic field, or doping level), we find hints of the 'dichotomy of anisotropy' in a wide range of quantum materials. Extended Data Table 3 lists the phase-shift ($\Delta\theta$) of various anisotropic physical quantities in materials including transition metal dichalcogenides (2H-NbSe$_2$[57], 4Hb-TaS$_2$[58]), Kagome superconductors[59], Moire superlattice[29] and doped topological insulator (Sr$_x$Bi$_2$Se$_3$[60], Cu$_x$Bi$_2$Se$_3$[61]) and cuprate superconductors $La_{2-x}Sr_xCuO_4$ with varying doping level[30].

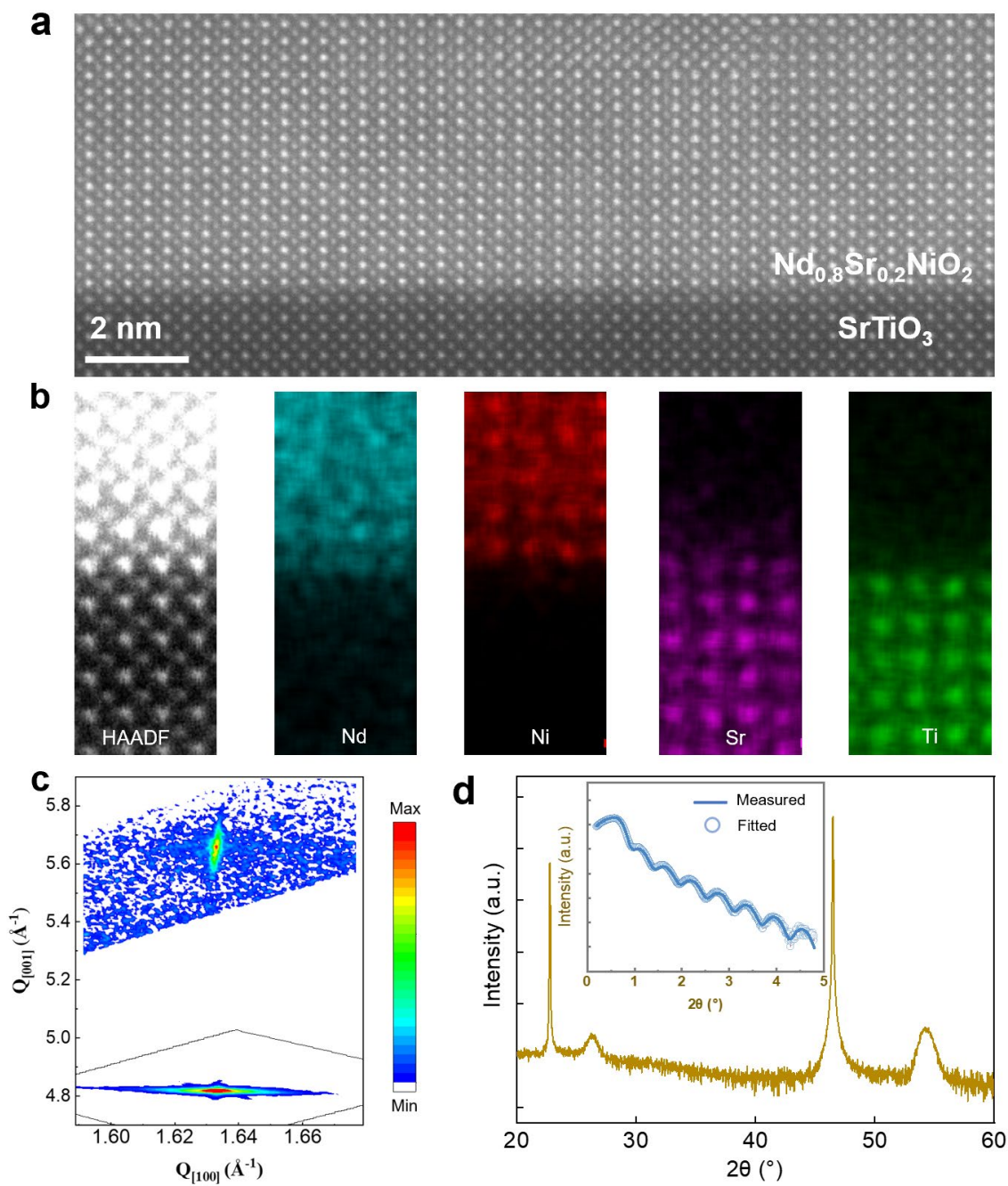

**Extended Data Fig 1| Structural characterization of Nd$_{0.8}$Sr$_{0.2}$NiO$_2$ thin film. a,** Cross-sectional high-angle annular dark-field scanning transmission electron microscopy (HAADF-STEM) image of a Nd$_{0.8}$Sr$_{0.2}$NiO$_2$/ SrTiO$_3$ interface. The scale bar is 2 nm. **b,** Zoom-in view of **a** and the corresponding energy-dispersive x-ray spectroscopy (EDS) maps with atomic resolution near the film-substrate interface. **c,** The reciprocal space maps (RSM) of superconducting Nd$_{0.8}$Sr$_{0.2}$NiO$_2$ films around the (103) SrTiO$_3$ diffraction peak. The RSM shows that the film and the substrate exhibit the same Q$_{[100]}$, indicating that our Nd$_{0.8}$Sr$_{0.2}$NiO$_2$ films are highly strained with the SrTiO$_3$ substrate. The dashed line indicates the nickelate film is fully in-plane constrained to the substrate. **d,** 2θ-ω x-ray diffraction spectrum of the nickelate film before and after topotactic reduction.

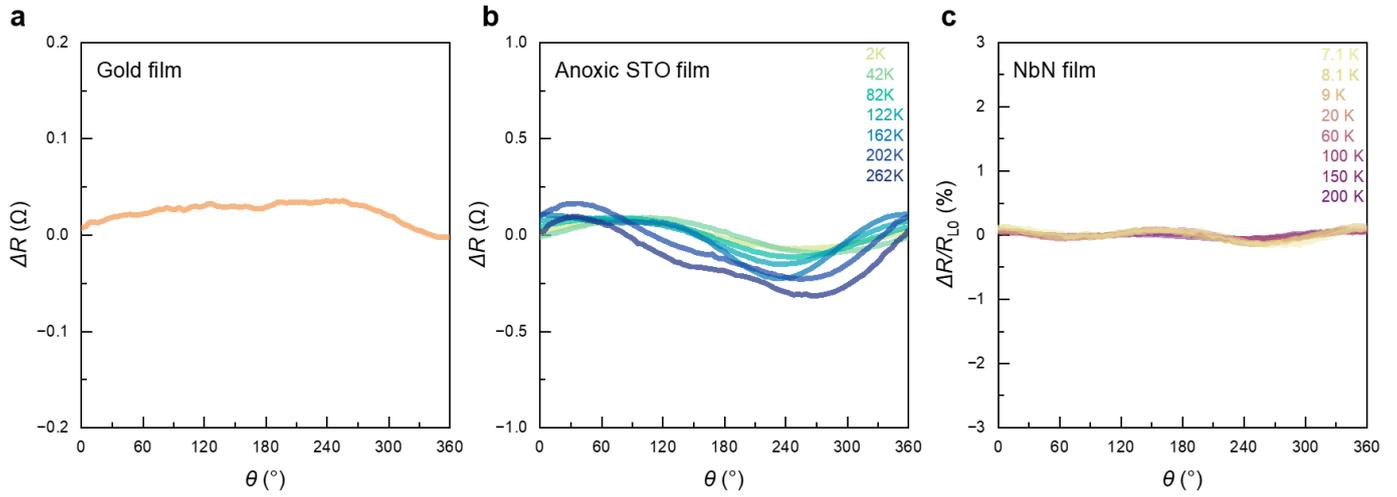

**Extended Data Fig 2 | Isotropic resistance in gold films, axonic SrTiO₃ films, and NbN films. a,** Angular resistance of gold film measured at 300 K. **b,** Angular resistance of anoxic SrTiO₃ film measured at selected temperature. **c,** Ratio of $\Delta R$ to background $R_{L,0}$ of NbN film measured at selected temperature.

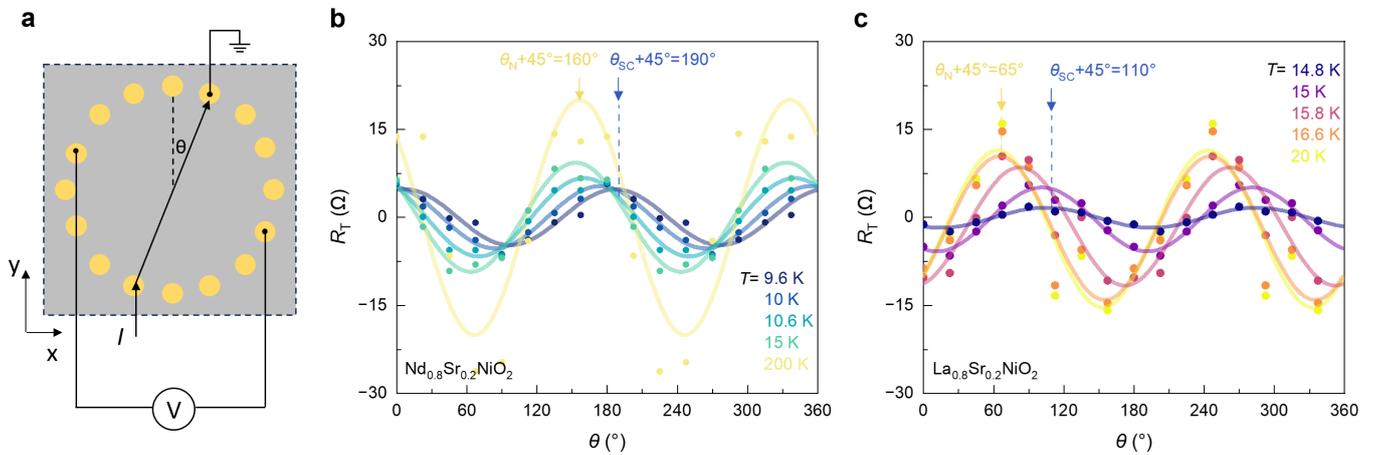

**Extended Data Fig 3 | Angular resistance measured for Nd₀.₈Sr₀.₂NiO₂, La₀.₈Sr₀.₂NiO₂ through a 'ring-shaped' device. a,** Schematic of measurement set-up. A ring of contact pads (yellow solid circles) deposited on the film (gray square) by evaporating gold through a shadow mask, without lithographic steps. The transverse resistance $R_T$ is measured using a set of four contacts in a cross. The angle-resolution of the ring-shaped device is 22.5°. **b,** angular transverse resistance of Nd₀.₈Sr₀.₂NiO₂ (NSNO-S2) at selected temperature, where $\theta_N$=115°, $\theta_{SC}$=155°. **c,** angular transverse resistance of La₀.₈Sr₀.₂NiO₂ (LSNO-S2) at selected temperature, where $\theta_N$=20°, $\theta_{SC}$=65°. Circles denote the experimentally-measured data points. Solid lines fit to $R_T(\theta) = \Delta R \sin[2(\theta - \theta_0)]$. The maximum of the $R_T(\theta)$ in normal state (near superconducting state) are marked by the yellow (blue) dashed lines.

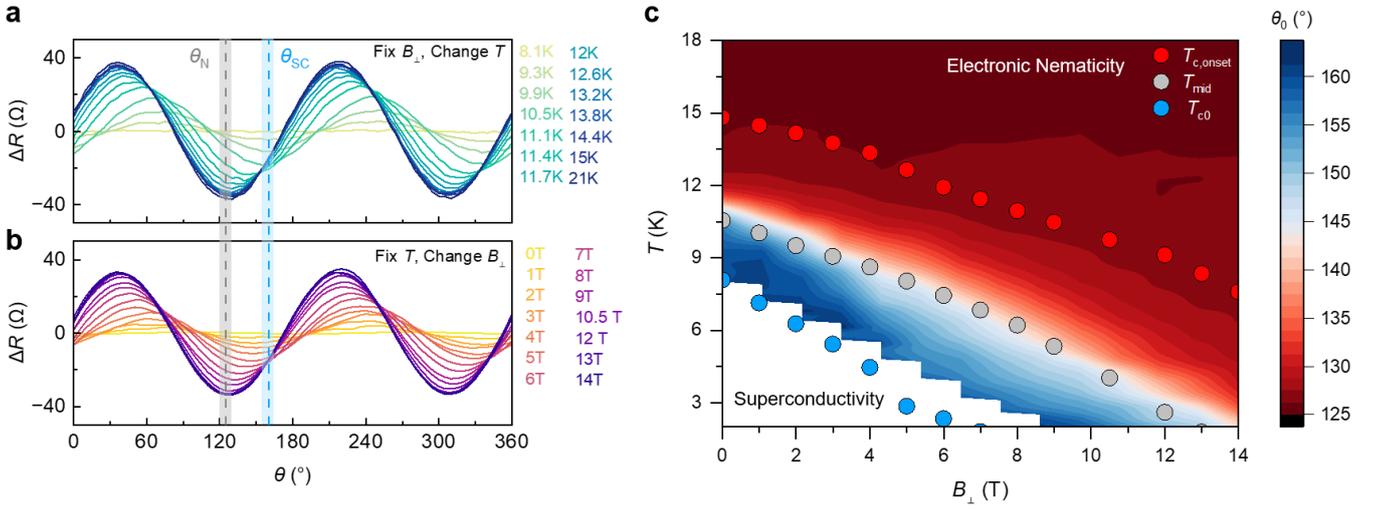

**Extended Data Fig 4| $T$ and $B_\perp$ dependence of $\Delta R_L(\theta)$ and the $B_\perp$-$T$ phase diagram for Nd$_{0.8}$Sr$_{0.2}$NiO$_2$. a,** Angle-dependent $\Delta R_L(\theta)$ at a series of temperature $T$ under zero field. **b,** Angle-dependent $\Delta R_L(\theta)$ measured under a series of perpendicular magnetic fields $B_\perp$ ranges from 0 T to 14 T, at fixed temperature $T$ =7.8 K. $\theta_0$ denotes the location of the minimum of $\Delta R_L(\theta)$. **c,** $T$ and $B_\perp$ dependence of $\theta_0$ for Nd$_{0.8}$Sr$_{0.2}$NiO$_2$ (NSNO-S1). Red dots denote the $T_c^{onset}$ at different $B_\perp$. Grey dots denote the $T_c$ at different $B_\perp$. Blue dots denote the $T_{c0}$ at different $B_\perp$. The definition for $T_c^{onset}$, $T_{mid}$, and $T_{c0}$ are the same as that in the Main text.

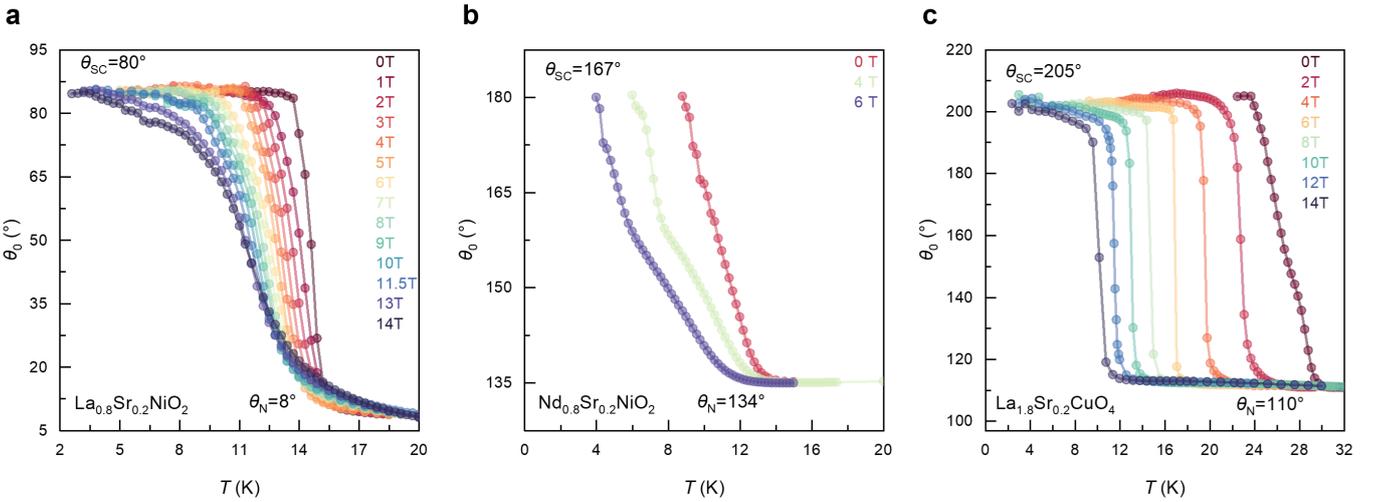

**Extended Data Fig 5| Shift of symmetric axes during the superconducting transition for La$_{0.8}$Sr$_{0.2}$NiO$_2$, Nd$_{0.8}$Sr$_{0.2}$NiO$_2$, and La$_{1.8}$Sr$_{0.2}$CuO$_4$ films. a,** Shift of symmetric axes of La$_{0.8}$Sr$_{0.2}$NiO$_2$ (LSNO-S2) with $\theta_N$ =8°, $\theta_{SC}$ =80°. **b,** Shift of symmetric axes of Nd$_{0.8}$Sr$_{0.2}$NiO$_2$ (NSNO-S3) with $\theta_N$ =134°, $\theta_{SC}$ =167°. **c,** Shift of symmetric axes of La$_{1.8}$Sr$_{0.2}$CuO$_4$ (LSCO-S1) with $\theta_N$=110°, $\theta_{SC}$=205°.

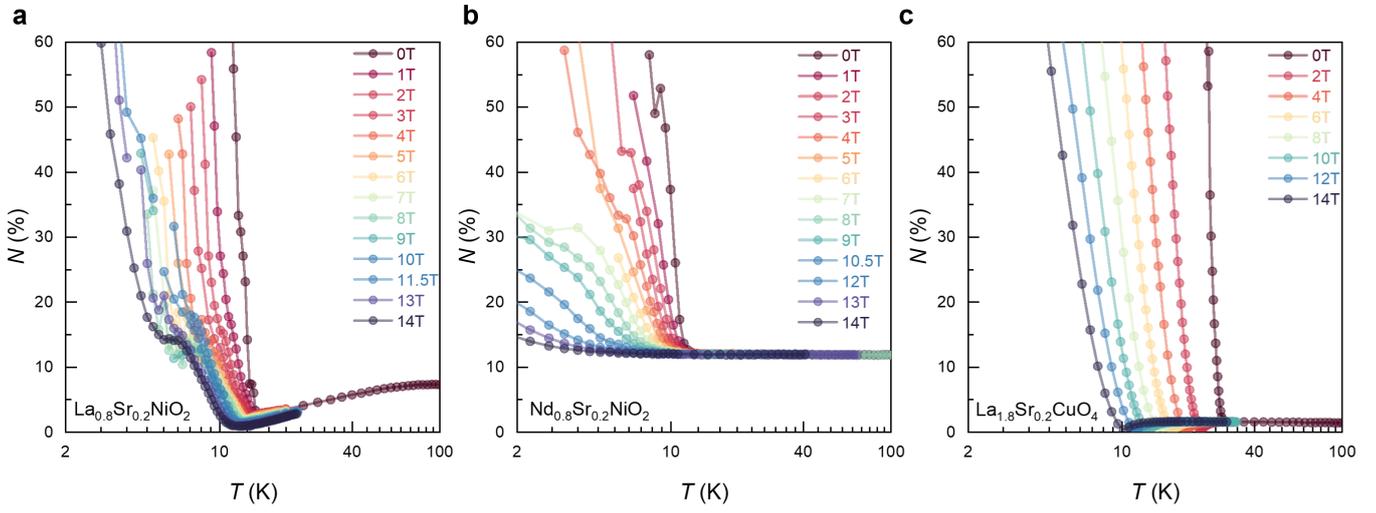

**Extended Data Fig 6| $T$ and $B$ dependence of strength of resistance anisotropy $N = 100\% * \Delta R/R_{L0}$ for Nd$_{0.8}$Sr$_{0.2}$NiO$_2$, La$_{0.8}$Sr$_{0.2}$NiO$_2$ and La$_{1.8}$Sr$_{0.2}$CuO$_4$ films. a,** La$_{0.8}$Sr$_{0.2}$NiO$_2$(LSNO-S2). **b,** Nd$_{0.8}$Sr$_{0.2}$NiO$_2$ (NSNO-S1). **c,** La$_{1.8}$Sr$_{0.2}$CuO$_4$ (LSCO-S1).

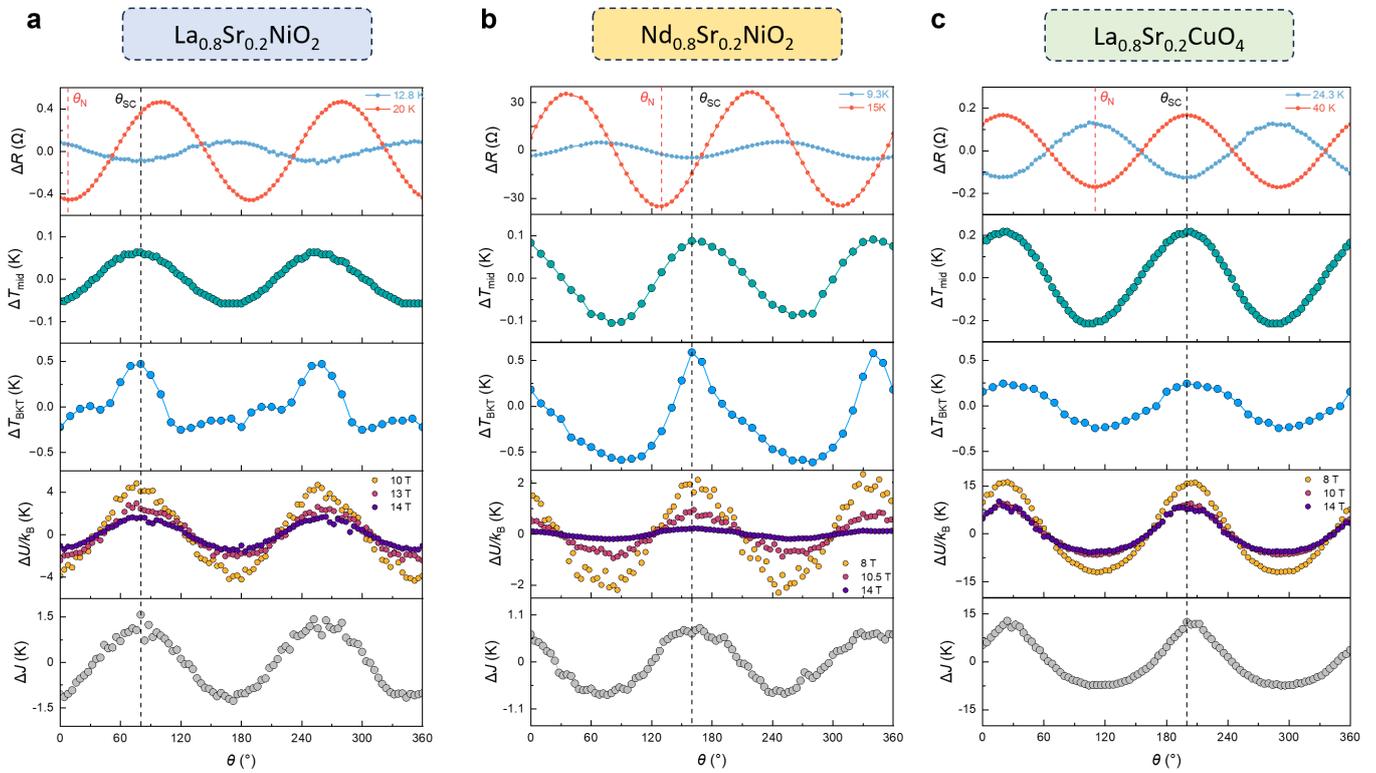

**Extended Data Fig 7| Consistency of physical quantities in La$_{0.8}$Sr$_{0.2}$NiO$_2$, Nd$_{0.8}$Sr$_{0.2}$NiO$_2$, and La$_{1.8}$Sr$_{0.2}$CuO$_4$.** The consistent alignment of symmetric axis of $\Delta R_L(\theta)$, $\Delta T_{mid}(\theta)$, $\Delta T_{BKT}(\theta)$, $\Delta U(\theta)/k_B$ and $\Delta J(\theta)$ in **a,** La$_{0.8}$Sr$_{0.2}$NiO$_2$ (LSNO-S1), **b,** Nd$_{0.8}$Sr$_{0.2}$NiO$_2$ (NSNO-S1), **c,** La$_{1.8}$Sr$_{0.2}$CuO$_4$ (LSCO-S1). Details for the La$_{0.8}$Sr$_{0.2}$NiO$_2$ and La$_{1.8}$Sr$_{0.2}$CuO$_4$ are shown and discussed in supplementary information.

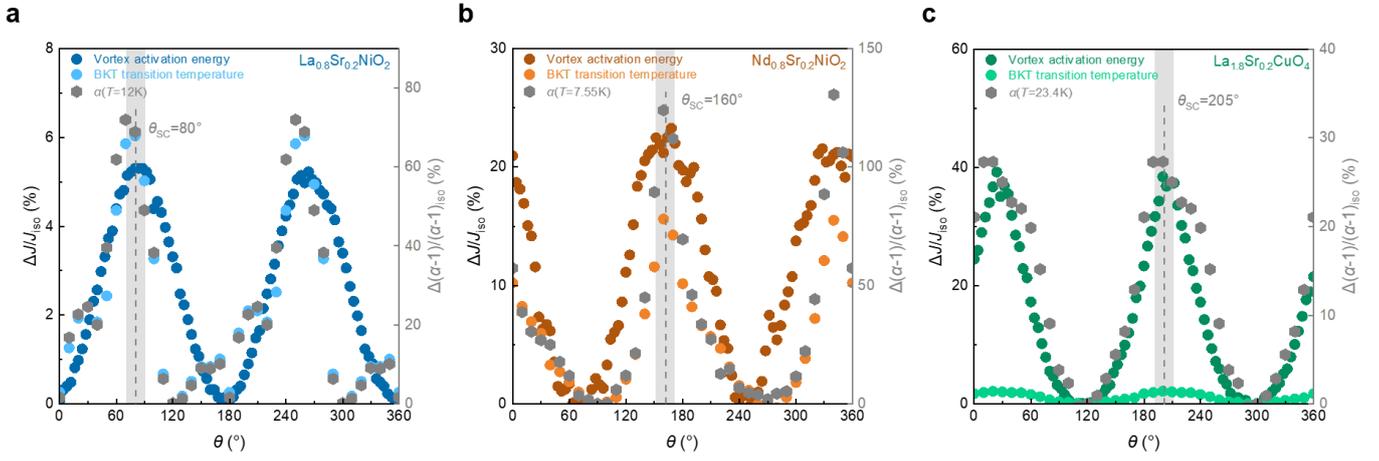

**Extended Data Fig 8| Consistency between the anisotropic superconducting phase stiffness calculated by different methods.** Circles show the anisotropy of $\Delta J/J_{iso}$ calculated by based on the anisotropic vortex activation energy, the normalizing procedure is shown before. Grey hexagons show the anisotropy of $\Delta(\alpha - 1)/(\alpha - 1)_{iso}$, where the $\alpha - 1$ is proportional to $J$. The consistent alignment of symmetric axis of $\Delta J/J_{iso}$ calculated by three different strategies is shown in **a,** La$_{0.8}$Sr$_{0.2}$NiO$_2$ (LSNO-S1), **b,** Nd$_{0.8}$Sr$_{0.2}$NiO$_2$ (NSNO-S1), **c,** La$_{1.8}$Sr$_{0.2}$CuO$_4$ (LSCO-S1). This consistency indicates the anisotropy of phase stiffness can be reproduced by two different methods.

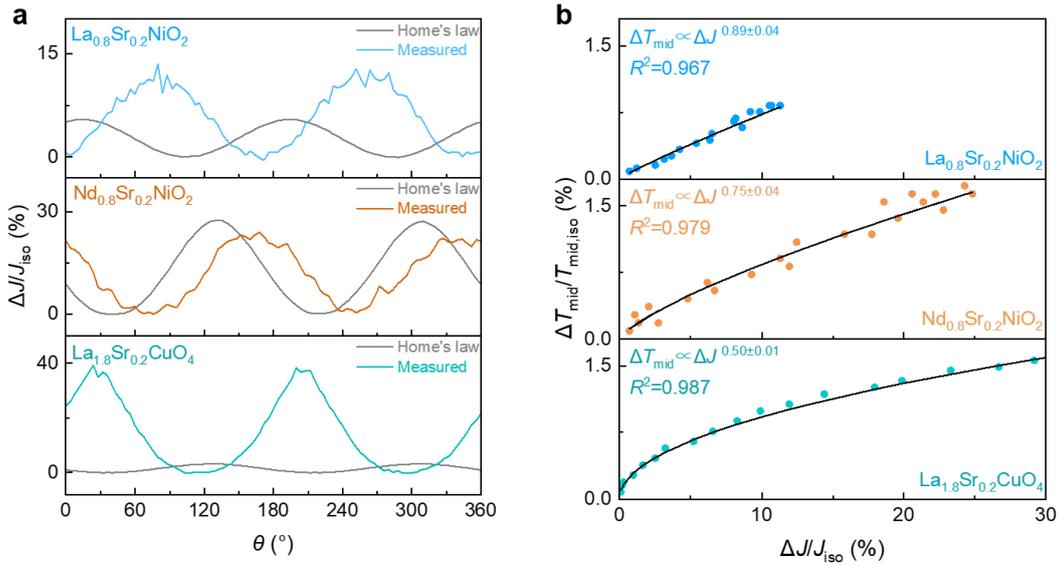

**Extended Data Fig 9| Home's law and sub-linear scaling in three materials. a,** Difference between the measured anisotropic $J(\theta)$ and that calculated by Home's law, $J(\theta) \propto \sigma_N(\theta)T_{mid}(\theta)$. The $J(\theta)$ is normalized with the same procedure shown in Fig. 4a-4c. Both the amplitude and the symmetric axis of the measured $J$ is irrelevant to that calculated from Home's law, demonstrating the anisotropy deviates from Home's law. **b,** The sub-linear scaling in three materials. The Coefficient of determination $R^2$ of the measured samples is 0.967 (La$_{0.8}$Sr$_{0.2}$NiO$_2$), 0.979 (Nd$_{0.8}$Sr$_{0.2}$NiO$_2$), 0.987 (La$_{1.8}$Sr$_{0.2}$CuO$_4$). The best-fitted index $n$ of all three materials is below 1, demonstrating the sub-linear scaling. Black line shows the best-fitted results.

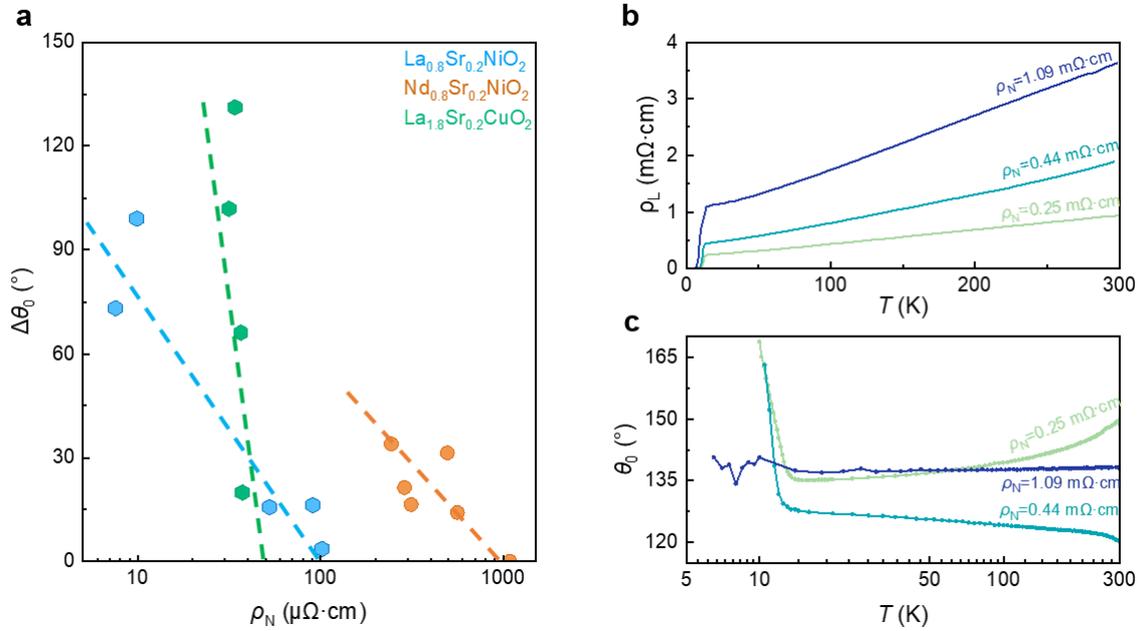

**Extended Data Fig 10| Relevance between the shift of symmetric axis $\theta_0$ ($\Delta\theta_0$) and disorder. a,** $\Delta\theta_0$ in samples with different degree of disorder. The $\Delta\theta_0$ is defined by the difference between the $\theta_0$ measured at $T_c^{onset}$ and $T_{c0}$. Normal state resistivity $\rho_N$ is the resistivity measured at $T_c^{onset}$. Larger $\rho_N$ indicates the larger degree of disorder. Dashed lines (guided by eyes) show a negative relevance between the shift of symmetric axis and degree of disorder in three materials. **b,** Temperature dependence of longitudinal resistivity $\rho_L(T)$ measured in $Nd_{0.8}Sr_{0.2}NiO_2$ samples. Lower resistivity indicates lower degree of disorder. **c,** Temperature dependence of symmetric axis $\theta_0$ measured in the corresponding samples shown in **b**. The absence of shift of $\theta_0$ during the superconducting transition in samples with large resistivity ($\rho_N$ =1.09 m$\Omega$·cm) indicates the quenched disorder potentially suppress this phenomenon. The shift of $\theta_0$ at higher temperature in relatively clean samples (green lines in **c**) is similar to the phenomena observed in $La_{1.8}Sr_{0.2}CuO_4$[2].

**Extended Data Table 1| Statistics of measured samples**

| Samples | $T_c^{onset}$ | $T_{c0}$ | $\theta_N$ | $\theta_{SC}$ | Source |
|---|---|---|---|---|---|
| $Nd_{0.8}Sr_{0.2}NiO_2$ (NSNO-S1) | 15 K | 8.7 K | 125° | 160° | Fig.2e |
| $Nd_{0.8}Sr_{0.2}NiO_2$ (NSNO-S2) | 15.6 K | 9 K | 115° | 155° | Extended Data Fig. 5b |
| $Nd_{0.8}Sr_{0.2}NiO_2$ (NSNO-S3) | 13 K | 10 K | 134° | 167° | Extended Data Fig. 3a |
| $La_{0.8}Sr_{0.2}NiO_2$ (LSNO-S1) | 16.1 K | 12.8 K | 8° | 80° | Extended Data Fig. 5c |
| $La_{0.8}Sr_{0.2}NiO_2$ (LSNO-S2) | 17.3 K | 14.5 K | 20° | 65° | Extended Data Fig. 3b |
| $La_{1.8}Sr_{0.2}CuO_4$ (LSCO-S1) | 32.1 K | 22 K | 110° | 200° | Extended Data Fig. 5c |
| $La_{1.79}Sr_{0.21}CuO_4$ (Bozovic's) | / | / | 120° | / | Table S1 in Ref [30] |

**Extended Data Table 2| Comparison between the calculated $J_{iso}$ and measured $J_0$**

| Samples | $T_{mid,iso}$ (K) | $\lambda_{iso}$ (nm) | $J_{iso}$ (K) | Reference | $T_c$ (K) | $\lambda_0$ (nm) | $J_0$ (K) |
|---|---|---|---|---|---|---|---|
| $La_{0.8}Sr_{0.2}CuO_4$ (LSCO-S1) | 26.5 | 269.3 | 51.3 | [9] | 26.8 | / | 52.8 |
| $Nd_{0.8}Sr_{0.2}NiO_2$ (NSNO-S1) | 11.0 | 377.4 | 6.6 | [62] | 17 | 750 | 1.75 |
| $La_{0.8}Sr_{0.2}NiO_2$ (LSNO-S1) | 14.2 | 219.5 | 20.4 | [62] | 10 | 1350 | 0.54 |

**Extended Data Table 3| Shift of physical quantities with external parameters**

| Materials | Tuning parameters | Physical quantity | $\Delta\theta$ (°) | Reference |
|---|---|---|---|---|
| 2H-NbSe$_2$ | $T$ | $B_{\parallel,c2}$ | 15 | [57] |
| | $B_{\parallel}$ | | | |
| 4H$_b$-TaS$_2$ | $T$ | $B_{\parallel,c2}$ | 20 | [58] |
| Magic Angle-Twist Bilayer Graphene | $B_{\parallel}$ | $B_{\parallel,c2}$ | 15 | [29] |
| | $n$ | $B_{\parallel,c2}$ | / | |
| Cu$_x$Bi$_2$Se$_3$ | $B_{\parallel}$ | $\Delta$ | 20.7 | [61] |
| Sr$_x$Bi$_2$Se$_3$ | $B_{\parallel}$ | $C$ | 32 | [60] |
| CsV$_3$Sb$_5$ | $B_{\parallel}$ | $B_{\parallel,c2}$ | 60 | [59] |
| La$_{1.96}$Sr$_{0.04}$CuO$_4$ | $T$ | $R_T$ | 2 | [30] |
| La$_{1.90}$Sr$_{0.10}$CuO$_4$ | | | -8 | |
| La$_{1.88}$Sr$_{0.12}$CuO$_4$ | | | 16 | |
| La$_{1.84}$Sr$_{0.16}$CuO$_4$ | | | 61 | |
| La$_{1.82}$Sr$_{0.18}$CuO$_4$ | | | 19 | |
| La$_{2-x}$Sr$_x$CuO$_4$ | $n$ | $R_T$ | / | [2] |

$T$ is the temperature, $B_{\parallel}$ is the in-plane magnetic field, $n$ is the doping level. $B_{\parallel,c2}$ denotes the in-plane critical field, $\Delta$ denotes the superconducting gap measured by Scanning Tunneling Microscopy, $C$ denotes the specific heat capacity, $R_T$ denotes the transverse resistance. The $\Delta\theta$ is defined by the maximum difference in the symmetry axis of the measured physical quantity when tuning the parameters. The $\Delta\theta$ for La$_{2-x}$Sr$_x$CuO$_4$ is calculated by $\theta(T = 295 \text{ K}) - \theta(T = T_c)$. The symmetric axis continuously shifts when varying doping level, rendering the $\Delta\theta$ is hard to quantified.